\documentclass[aps,prl,twocolumn,showpacs,preprintnumbers,amsmath,amssymb]{revtex4-1}

\newcommand{\be}{\begin{equation}}
\newcommand{\ee}{\end{equation}}
\newcommand{\bea}{\begin{eqnarray}}
\newcommand{\eea}{\end{eqnarray}}

\newcommand{\mb}{\mathbf}

\usepackage{graphicx}
\usepackage{dcolumn}
\usepackage{bm}

\usepackage{color}

\begin{document}
\title{Antiferromagnetic Order in a Semiconductor Quantum Well with Spin-Orbit Coupling}
\author{D. C. Marinescu}
\affiliation{Department of Physics and Astronomy, Clemson University,
Clemson, South Carolina 29634, USA}

\date{\today}
\begin{abstract}
An argument is made on the existence of a low-temperature itinerant antiferromagnetic (AF) spin alignment, rather than persistent helical (PH), in the ground state of a two dimensional electron gas in a semiconductor quantum well with linear spin-orbit Rashba-Dresselhaus interaction at equal coupling strengths, $\alpha$. This result is obtained on account of the opposite-spin single-particle state degeneracy at $\mb k = 0$ that makes the spin instability possible. A theory of the resulting magnetic phase is formulated within the Hartree-Fock approximation of the Coulomb interaction. In the AF state the direction of the fractional polarization is obtained to be aligned along the displacement vector of the single-particle states.
\end{abstract}
\pacs{72.25.-b, 72.10.-d}
\maketitle

\begin{figure}
 \begin{center}
    {\includegraphics[width=3in]{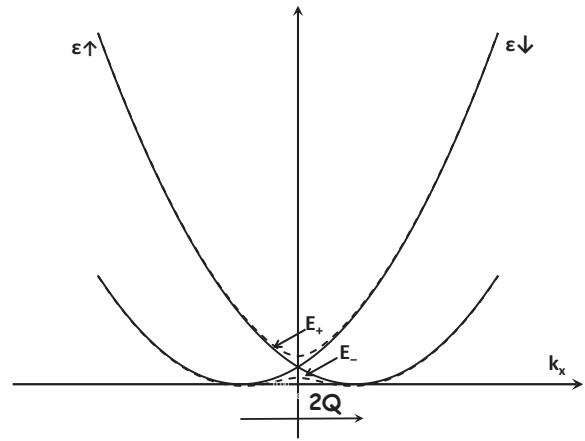}}
  \end{center}
 \caption{\small The single particle spectrum of a 2D electron system with equal Rashba-Dresselhaus linear couplings. The opposite-spin energies $\varepsilon_\uparrow$ and $\varepsilon_\downarrow$ are degenerate at $\mb k = 0$, the source of a spin instability that leads to the formation of two new quasiparticles of energies $E_-$ and $E_+$. The states that overlap at $\mb k = 0$ are plane waves of the same momentum.}
 \label{fig1}
\end{figure}

Driven by the desire to control the motion of the electron spin by electric means, the ten-year old resurgence of interest in understanding the consequences of the spin-orbit interaction (SOI) in two dimensional (2D) electron systems has produced a significant body of work centered on the coupling between the electron spin and its momentum embodied by the Rashba (R) \cite{rashba} and Dresselhaus (D) \cite{dresselhaus} interactions. They both appear in zinc-blende semiconductor structures as a result of an inversion asymmetry, be it in the structure (R) or in bulk (D). Among the many interesting phenomena discovered by analyzing the effects of SOI in quantum wells,
the prediction of the persistent helical state (PHS) \cite{bernevig06} received widespread attention since it indicated the existence of a special case of spin symmetry. The essence of the argument supporting the emergence of PHS is that at equal Rashba-Dresselhaus strengths $\alpha$,  the displacement in the momentum space of the opposite-spin single particle energies, indexed by momentum $\hbar \mb k$ as shown in Fig.~\ref{fig1}, by a wave vector $2Q =4m^*\alpha/\hbar \hat{\mb x}$ ($m^*$ is the electron mass) leads to the energy-invariance condition
\be
\varepsilon_{\uparrow}(\mb k + 2\mb Q) = \varepsilon_\downarrow(\mb k)\;,\label{eq:invariance}
\ee
which would favor the same-energy spin-flip processes that enable the real space formation of a spin helix. The created helical spin density wave is characterized by a real-space spatial rotation of the spin polarization periodic in space with period $2\mb Q\cdot \mb r$.
This spiral spin symmetry was predicted to be robust under all forms of spin-independent scattering, including Coulomb interaction.

Ulterior experimental evidence seemed to support this conclusion by testing several characteristic features of this state, such as the existence of very long spin relaxation rates \cite{koralek09, walser12} and the absence of the antilocalization correction to the conductivity \cite{kohda12}. Although it was found that the cubic Dresselhaus term can act as a perturbation \cite{koralek09,zhou}, this effect is in general considered to be small. For the purpose of the argument put forward in this paper, it is important to note that these experimental results are not incompatible with an itinerant antiferromagnetic arrangement. Moreover, long spin relaxation times and the absence of the antilocalization corrections were both previously discussed in contexts that did not involve PHS \cite{egues,pikus}.

The focus of this letter is not the energy invariance condition, Eq.~(\ref{eq:invariance}), but rather the degeneracy of the opposite-spin single-particle states at $\mb k = 0$, $\varepsilon_{\uparrow}(0) = \varepsilon_{\downarrow}(0)$, as seen in Fig.~\ref{fig1}.  By following the traditional analysis of spin instabilities in Fermi systems \cite{awo60,awo62,gfgbook}, I show that this is the real source of a potential long-range magnetic order. Since this is a many-body system, the existence of such an instability in the single-particle spectrum is not enough, by itself, to justify the establishment of such a collective arrangement. Therefore, I use the Hartree-Fock approximation of the Coulomb interaction to obtain the single-particle energy states and self-consistently derive the fundamental gap equation of the system which is solved in a simplifying picture, an algorithm whose validity in the case of simple Fermi liquids is supported by extensive literature \cite{gfgbook}. Our results indicate that at low temperature an antiferromagnetic (AF) state is possible, whereas the trivial paramagnetic phase is likely to prevail at higher temperatures.

Degeneracies between opposite-spin states in the single particle spectrum have been previously considered in many instances. In the case of simple Fermi liquids they were first suggested by Overhauser who demonstrated that, within the Hartree-Fock (HF) approximation, Slater determinants constructed out of linear combinations of electron states with variable spin orientation in $\mb k$-space will generate a total energy lower than that of the paramagnetic configuration \cite{awo60,awo62}. Allowing for the rotation of the spin, however, required the removal of the fundamental spin degeneracy in the single electron states of momentum $\mb k$ for the two spin orientations, $\varepsilon_{\uparrow}(\mb k) = \varepsilon_{\downarrow}(\mb k)$ by artificially introducing a momentum displacement $2\mb Q$ in $\mb k$-space. Since the new quasiparticle eigenstates are obtained from the linear superposition of plane waves of momenta $\mb k$ and $\mb k + 2\mb Q$, expectation values of the spin operators produced a spiral spin-density-wave (SDW) \cite{awo60} characterized by a spin polarization whose direction was rotating in the real space with a periodicity given by $2\mb Q\cdot \mb r$.

Spin instabilities were also obtained in the presence of a magnetic field in Hall systems. Such driven magnetic phases were studied both theoretically and observed experimentally in single quantum wells \cite{gfg85prb,fang}, double layers \cite{zheng97,dassarma98,pellegrini}, and in multilayers \cite{brey,marinescu00,marinescu11}. Depending on the nature of the single particle states involved in the creation of the spin-degeneracy, the resulting magnetic phases where found to be either spiral spin density waves (SDW) \cite{marinescu00, marinescu11} or canted antiferromagnetic \cite{zheng97, dassarma98}.

In contrast to the classic SDW case discussed above, in the 2D electron gas with SOI, the opposite-spin states involved in the degeneracy are {\it naturally} separated in the momentum space by the Rashba interaction, leading to a coupling that has to necessarily occur between plane waves of the same momentum $\mb k$. Consequently, as we show below, the spin polarization of each quasiparticle state loses its real space variation, and rather maintains a fixed direction in space, parallel with the direction of the displacement vector $\mb Q$.

The basic physical system for our calculation is a 2D electron system in a semiconductor quantum well with Rashba-Dresselhaus spin-orbit of strengths $\alpha$ and $\beta$ respectively placed in the $\hat{\mb x}-\hat{\mb z}$ plane. Throughout this calculation we neglect the cubic Dresselhaus term. A positive background is understood to assure charge neutrality.
The single particle Hamiltonian of an electron of wave-vector $\mb k = \{k_x,k_z\}$ and spin $\mb \sigma = \left\{\sigma_x,\sigma_y,\sigma_z\right\}$ is given by,
\be
H_1 = \frac{\hbar^2k^2}{2m^*} + \hbar\alpha\left(\sigma_zk_x - \sigma_xk_z\right) + \hbar\beta(\sigma_xk_x - \sigma_zk_z)\;.\label{eq:H1}
\ee
In the particular case $\alpha = \beta$ the Hamiltonian is written,
\be
H_1 = \frac{\hbar^2k^2}{2m^*} + \hbar\alpha (\sigma_x+\sigma_z)(k_z-k_x)\;,\label{eq:H2}
\ee
suggesting that a simultaneous unitary transformation in the spin space to ${\sigma}_z' = (\sigma_x+\sigma_z)/\sqrt{2}$ and a rotation in real space to $k'_x = (k_x-k_z)/\sqrt{2}; k'_z =(k_z+k_x)/\sqrt{2} $ leads to the simpler form recognized in Ref.~\cite{bernevig06},
\bea
H_1 & = & \frac{\hbar^2( k'_x)^2}{2m^*}+\frac{\hbar^2( k'_z)^2}{2m^*} - 2\hbar\alpha k'_x\sigma_z' \nonumber \\
& = & \frac{\hbar^2\left(k'_x- Q\sigma_z'\right)^2}{2m^*}+\frac{\hbar^2(k')_z^2}{2m^*}-\frac{\hbar^2Q^2}{2m^*}\;,\label{eq:H3}
\eea
where $Q = 2m^*\alpha/\hbar$.
Henceforth I will refer to Eq.~(\ref{eq:H3}) as the fundamental single particle Hamiltonian of the problem and consequently, for ease of notation, drop the prime indices from the momenta and spins.

For a sample of unit area, the spin-dependent eigenstates of Eq.~(\ref{eq:H3}) are plane waves,
\be
\psi_{\mb k, \sigma}(\mb r) = e^{i\mb k \cdot \mb r}|\sigma\rangle\;,\label{eq:eigenstate}
\ee
of energy eigenvalues, written in respect with the constant $-{\hbar^2Q^2}/{2m^*}$,
\be
\varepsilon_{\mb k,\sigma} = \frac{\hbar^2(k_x - \sigma\mb Q)^2}{2m^*} + \frac{\hbar^2k_z^2}{2m^*}\;,\label{eq:eigenvalue}
\ee
where $\sigma$ is  $1$ for $|\uparrow\rangle$ and $-1$ for $|\downarrow\rangle$.

The energies (\ref{eq:eigenvalue}) satisfy the invariance expressed in Eq.~(\ref{eq:invariance}), a symmetry property exploited in the argument that the total lowering and raising spin operators, $S_{\mb Q} = \sum_{\mb k}c_{\mb k,\downarrow}^\dag c_{\mb k + 2\mb Q,\uparrow}$ and $S_{\mb Q}^+ = \sum_{\mb k} c_{\mb k + \mb Q,\uparrow}^\dag c_{\mb k,\downarrow}$, respectively, commute with the kinetic energy part of the total Hamiltonian of the system $\sum_{\mb k}\varepsilon_{\mb k}c_{\mb k,\sigma}^\dag c_{\mb k,\sigma}$ ($c_{\mb k}$ and $c_{\mb k + \mb Q}^\dag$ represent single state electron operators associated with the states (\ref{eq:eigenstate})). The appearance of the persistent helical state, as discussed in Ref.~\cite{bernevig06}, is the result of the same-energy spin-flips between the states associated with the energy degeneracy in Eq.~(\ref{eq:invariance}), enabling the existence  of non-zero matrix elements for $S_{\mb Q}^\dag$ and $S_{\mb Q}$ and thus generating the spatial dependence of the spin polarization which is periodic with $2\mb Q \cdot \mb r$. As we show in detail below, non-zero matrix elements of the total raising and lowering spin operators, $S_{\mb Q}^\dag$ and $S_{\mb Q}$, can occur only between Slater determinants constructed out of single particle states that are linear combinations of $\psi_{\mb k, \uparrow}$ and $\psi_{\mb k,\downarrow}$, a situation that violates the minimum energy principle for the total energy of the system. Moreover, in the limit $\mb Q = 0$ which describes the usual paramagnetic system, the same arguments hold leading to the un-physical result of a non-zero off-diagonal spin polarization.

In other words, the energy configuration based on the invariance in Eq.~(\ref{eq:invariance}) and the superposition of opposite spin states Eq.~(\ref{eq:eigenstate}), necessary for an off-diagonal spin polarization to result, is higher than that of the paramagnetic state and should be unstable, exactly in the same way in which linear combinations of single-particle, opposite-spin states in the usual, completely degenerate Fermi liquid (without spin-orbit interaction) lead to a higher total energy than the paramagnetic state and are, thus, never unrealized.

In the following considerations, I focus instead on another important feature of this spectrum, namely  the accidental degeneracy that happens at $\mb k = 0$, when the opposite spin energies become equal. As argued in the introduction, extensive literature supports the idea that on account of this degeneracy, a no-kinetic energy cost spin instability can occur with consequent broken spin symmetry magnetic ordering if it produces a lower energy than the paramagnetic state.
The problem needs to be studied in the presence of the Coulomb interaction, since changes in the spin structure involve energy costs of the whole many-body system. It has been long known that the introduction of the Coulomb interaction in the total energy of an interacting electron system has the merit of decreasing the total potential energy of the system, by enhancing the negative exchange contribution that favors parallel spin alignment. Simultaneously, however, the increase in the kinetic energy on account of the Pauli exclusion principle, naturally prevents an all ferromagnetic alignment. It is in this situation that the original model for the spin-density wave formation was proposed by Overhauser in Refs.~\cite{awo60,awo62} as a lower total energy alternative to the paramagnetic alignment, by allowing for the continuous rotation of the electron spin in the momentum space.

We consider therefore the many-body Hamiltonian of the system written in terms of the single particle eigenstates identified in Eq.~(\ref{eq:eigenstate}) represented by the creation and destruction operators $c^\dag_{\mb k,\sigma}, c_{\mb k,\sigma}$.
The equilibrium many-body Hamiltonian is composed of a kinetic energy part and the usual Coulomb repulsion,
written as
\bea
H& = & \sum_{\mb k,\sigma}\varepsilon_{\mb k,\sigma}c_{\mb k,\sigma}^\dagger c_{\mb k,\sigma}\nonumber\\
 & + & \frac{1}{2}\sum_{\substack{\mb k,\mb q, \mb k'\\\sigma,\sigma'}}v(q)
c_{\mb k + \mb q,\sigma}^\dag c_{\mb k',\sigma'}^\dag c_{\mb k'+\mb q,\sigma'}c_{\mb k,\sigma}\;. \label{eq:many}
\end{eqnarray}
$v(\mb q) = 2\pi e^2/q$ is the Coulomb interaction matrix element in two dimensions.
The ground state energy of the system, is obtained by averaging of the
total Hamiltonian on the ground state wave function, a process that involves certain approximations of the
interaction terms. In Hartree-Fock, the ground-state average
 of the
product of four operators is factorized into a product of two
two-particle operators \cite{gfgbook}. For now, no
assumption is made on the nature of the ground state. Thus, with
$<\ldots,\ldots>$ representing the ground state average, the
interaction becomes,
\begin{eqnarray}
& & \langle H_{int} \rangle = \frac{1}{2}
\sum_{\substack{\mb k,\mb q\\\sigma,\sigma'}}{v}(q)\left[\langle c_{\mb k + \mb q,\sigma}^{\dag}
c_{\mb k,\sigma}\rangle \langle c_{\mb k\sigma'}^{\dag}
c_{\mb k' + \mb q,\sigma^{'}}\rangle \right.\nonumber \\
& - &
\left.\langle c_{\mb k + \mb q,\sigma}^{\dag}
c_{\mb k' + \mb q,\sigma'}\rangle \langle c_{\mb k',\sigma'}^{\dag}
c_{\mb k,\sigma}\rangle \right]\;.\label{eq:interaction}
\end{eqnarray}

In the first term of Eq.~(\ref{eq:interaction}) one recognizes the direct
interaction, $\langle c_{\mb k + \mb q,\sigma}^{\dag}
c_{\mb k,\sigma}\rangle \langle c_{\mb k\sigma'}^{\dag}
c_{\mb k' + \mb q,\sigma^{'}}\rangle =n_{\mb k,\sigma}^0n_{\mb k + \mb q}^0\delta_{\mb q,0}$ , which is canceled out  by
the positive background ($n_{\mb k,\sigma}^0$ is the ground state occupation number of the single particle state). The second term represents the exchange interaction.
Since in this formalism, the ground state averages are evaluated on a state that in itself is not known, apriori assumptions on the magnetic ordering of the ground state need to be made. A ground state that assumes only a parallel spin alignment, be it paramagnetic or ferromagnetic, will generate non-zero averages in $\langle c_{\mb k',\sigma'}^{\dag}
c_{\mb k,\sigma}\rangle$ only for $\sigma = \sigma'$, which is the usual exchange in Fermi systems with paramagnetic or ferromagnetic configurations. If it is hypothesized, however, that the ground state is based on a non-parallel spin alignment, then by default, one needs to assume that a matrix element of the type $\langle c_{\mb k,\sigma}^{\dag}
c_{\mb k',\sigma'}\rangle \neq 0$ even for $\sigma \neq \sigma'$. This is the fundamental paradigm of the SDW formation in simple metals \cite{awo60}, an assumption also pursued here, such that we write,
\begin{eqnarray}
& & \langle H_{int}\rangle =  -\frac{1}{2}\sum_{\mb k,\mb q,\sigma}
{v}(q)\langle c_{\mb k + \mb q,\sigma}^{\dag}
c_{\mb k + \mb q,\sigma}\rangle
\langle c_{\mb k,\sigma}^{\dag} c_{\mb k,\sigma} \rangle\nonumber\\
& - & \frac{1}{2}\sum_{\mb k,\mb q}
{v}(q)\langle c_{\mb k + \mb q,\sigma}^{\dag}
c_{\mb k + \mb q,\bar{\sigma}}\rangle \langle c_{\mb k,\bar{\sigma}}^{\dag}
c_{\mb k,\sigma}\rangle \;,\label{eq:int}
\end{eqnarray}
where $\bar{\sigma}$ is the opposite of $\sigma$. As we show below, the second term in Eq.~(\ref{eq:int}) is responsible for the antiferromagnetic coupling of the electron spins.

The total Hartree-Fock Hamiltonian of the system is then liniarized by means of a canonical
Bogoliubov-Valatin (BV) transformation \cite{gfgbook}. This introduces
two new fermionic operators $u _{\mb k }$ and $ v_{\mb k }$ ,
\begin{eqnarray}
u_{\mb k}&=&\cos \theta_{k_x} c_{\mb k,\uparrow} + \sin \theta_{k_x} c_{\mb k,\downarrow}\;, \nonumber \\
v_{\mb k}&=&-\sin \theta_{k_x}c_{\mb k,\uparrow}+\cos
\theta_{k_x} c_{\mb k,\downarrow}\;.\label{eq:bv}
\end{eqnarray}
which are continuous functions of the spin inclination angle $\theta_{k_x}$, the variational parameter of the
transformation. In this form, it is transparent that $u_{\mb k}$ and $v_{\mb k}$ describe electron states whose spin composition varies continuously throughout the momentum space. The sole $k_x$ dependence of $\theta$ is justified by the spectrum presented in Fig.~\ref{fig1}. The electron operators that enter these expressions, $c_{\mb k,\sigma}$ have to be those of the opposite-spin states that are involved in the accidental spin degeneracy, i.e., the spin rotation is allowed only if there is no kinetic energy cost. In the case of the Rashba system these are states of the same index $\mb k$, a different situation than that in Refs. \cite{awo60,awo62} where the states involved corresponded to $\mb k$ and $\mb k + \mb Q$. In real space, the state functions associated with $u_{\mb k}$ and $v_{\mb k}$ are written as linear superpositions of the single particle eigenstates at the point of degeneracy, Eq.~(\ref{eq:eigenstate}),
\bea
\Psi_{\mb k+} = \left(\cos \theta_{k_x}|\uparrow\rangle + \sin \theta_{k_x}|\downarrow\right)e^{i\mb k \cdot \mb r}\;,\nonumber\\
\Psi_{\mb k-} = \left(-\sin \theta_{k_x}|\uparrow\rangle + \cos \theta_{k_x}|\downarrow\right)e^{i\mb k \cdot \mb r}\;. \label{eq:states}
\eea
Immediately, one observes that the spin polarization associated with these single-particle states has only $x$ and $z$ components, given by
\bea
\mb P_\pm & = & \hat{\mb x}\langle \Psi_{\mb k,\pm}|\mb \sigma_{x}|\Psi_{\mb k,\pm}\rangle + \hat{\mb z}\langle \Psi_{\mb k,\pm}|\mb \sigma_{z}|\Psi_{\mb k,\pm}\rangle \nonumber\\
& = & \pm \hat{\mb x}\sin 2\theta_{k_x} \pm \hat{\mb z}\cos 2\theta_{k_x}\;,
\eea
missing any real space variation.

The substitution of the electron operators by the
Eqs.~(\ref{eq:bv}) leads to an expression for the ground state
energy that depends on averages of the newly introduced operators,
$u_{\mb k}$ and $v_{\mb k}$. There are four types of terms that
appear. Two represent the same particle averages,
$<u_{\mb k}^\dag u_{\mb k}>$ and
$<v_{\mb k}^\dag v_{\mb k}>$, and two mixed-ones,
$<u_{\mb k}^\dag v_{\mb k}>$ and
$<v_{\mb k}^\dag u_{\mb k}>$. The first category can be easily
associated with the occupation numbers of two new quasiparticles,
while the second describes the excitation processes of
these quasiparticles. Since the ground state of the system is of
interest, we will neglect the quasiparticle excitations. Thus, by
the means of the BV transformation, the system of interacting
electrons is transformed into a system of non-interacting
quasiparticles.

As a function of the quasiparticle occupation numbers, $f_{\mb k,+} =
<u_{\mb k}^\dag u_{\mb k}>$ and $f_{\mb k,-} = <v_{\mb k}^\dag
v_{\mb k}>$, the ground state energy becomes,
\begin{eqnarray}
& &\langle H \rangle_{HF}
=\sum_{\mb k}\tilde{\varepsilon}_{\mb k,\uparrow}\left(\cos^{2}
\theta_{{k_x}} f_{\mb k,+}  + \sin^2\theta_{k_x}f_{\mb k,-}\right)\nonumber\\
&+& \sum_{{\mb k}}\tilde{\varepsilon}_{\mb k,\downarrow}\left(\sin^{2}
\theta_{k_x}f_{\mb k,+}+\cos^2\theta_{k_x} f_{\mb k,-}\right)
\nonumber\\
&-&\frac{1}{4}\sum_{{\mb k},{\mb k'}}{v}(\mb k'-\mb k)\sin
2\theta_{k_x}\sin 2
\theta_{k_x'}(f_{\mb k,+}-f_{\mb k,-})\nonumber \\
& \times & (f_{\mb k',+}-f_{\mb k',-})\;,\nonumber\\
\label{eq:sdw-energy}
\end{eqnarray}
where we introduced $\mb q = \mb k'-\mb k$ and the single-particle energies in the HF approximation,
\begin{equation}
\tilde{\varepsilon}_{\mb k,\uparrow} =\varepsilon_{\mb k,\uparrow} -
\sum_{\mb k'}v(\mb k'-\mb k)\left (\cos^2
\theta_{k_x'}f_{\mb k',+} + \sin^2\theta_{k_x'}f_{\mb k',-}\right) \;,
\end{equation}
and
\begin{equation}
\tilde{\varepsilon}_{\mb k,\downarrow} =
\varepsilon_{\mb k,\downarrow}-
 \sum_{\mb k'}{v}(\mb k'-\mb k) \left (\sin^2
\theta_{k_x'}^{2}f_{\mb k',+} + \cos^2\theta_{k_x'}f_{\mb k',-}\right)\;.
\end{equation}

We note that the canonical transformation, Eq.~(\ref{eq:bv}) preserves the total number of particles since
\be
N = \sum_{\mb k,\sigma} c_{\mb k,\sigma}^\dag c_{\mb k,\sigma} = \sum_{\mb k,i=\pm} f_{\mb k,i}\;.
\ee
At finite temperatures, a fermionic entropy term can be added to the ground state energy,
\be
S = - k_B\sum_{\mb k,i=\pm}\left[f_{\mb k,i} \ln f_{\mb k,i} + (1-f_{\mb k,i})\ln (1-f_{\mb k,i})\right]\;.
\ee

By minimizing the grand canonical thermodynamic function, written for a chemical potential $\mu$,
\be
\Omega(T,V,\mu) = \langle H \rangle_{HF} - \mu N - TS\;,
\ee
in respect with $\theta_{\mb k}$ and the two quasiparticle occupation numbers $f_{1\mb k}$ and $f_{2\mb k}$, a set of three coupled self-consistent equations are obtained. First, by minimizing in respect with $\theta_{k_x}$ one obtains the gap equation,
\begin{equation}
\tan\left(2\theta_{k_x}\right)=
\frac{g(\mb k)}{\tilde{\varepsilon}_{\mb k, \downarrow}-\tilde{\varepsilon}_{\mb k, \uparrow}}
\;, \label{eq:gap}
\end{equation}
where the antiferromagnetic gap is
\begin{equation}
g_{\mb k} =
\sum_{\mb k'}v(\mb k'-\mb k)\sin2\theta_{k_x'}\;,\label{eq:sl-gap}
\end{equation}
Eq.~(\ref{eq:gap}) is a non-local, self-consistent expression, since
the solution is dependent on the values of the inclination angle
throughout the Brillouin zone. By minimizing $\Omega(T,A,\mu)$ in respect with the two occupation numbers, the single quasiparticle energies are obtained
as
\begin{equation}
 E_{\pm}(\mb k) =\frac{1}{2} \left[
\tilde{\varepsilon}_{\mb k,\downarrow}+\tilde{\varepsilon}_{\mb k,\uparrow}
\mp
\sqrt{(\tilde{\varepsilon}_{\mb k,\downarrow}-\tilde{\varepsilon}_{\mb k,\uparrow})^{2}+g_{\mb k}^{2}}\right] + \frac{\hbar^2k_z^2}{2m^*}
\;.
\end{equation}
As depicted by dashed lines in Fig.~\ref{fig1}, $E_{+}$ and $E_-$ are separated at $k_x = 0$ by the $g_{\mb k}$, the antiferromagnetic gap.

At $T = 0$K, when it can be assumed that only the lowest quasiparticle level is occupied, i.e. $f_{\mb k,+} = 1$ and $f_{\mb k,-} = 0$, the stability condition for the antiferromagnetic phase is $\partial^2
<H>_{HF}/\partial \theta^2_{k_x} <0$, which is always realized when
a solution to the gap equation is found since,
\begin{equation}
\frac{\partial^2 <H>_{HF}}{\partial \theta^2_{k_x}} =
-\sqrt{(\tilde{\varepsilon}_{\mb k,\downarrow}-\tilde{\varepsilon}_{\mb k,\uparrow})^{2}+g_{\mb k}^{2}}\;.
\end{equation}

In the same temperature regime, one can combine Eqs.~(\ref{eq:gap}) and (\ref{eq:sl-gap}) in a single expression,
\be
g_{\mb k} = \sum_{\mb k'}v(\mb k-\mb k')\frac{g_{\mb k'}}{\sqrt{(\tilde{\varepsilon}_{\mb k}-\tilde{\varepsilon}_{\mb k'})^2 + g_{\mb k'}^2}}\;.\label{eq:gap3}
\ee
The trivial solution of Eqs.~(\ref{eq:gap}) and (\ref{eq:gap3}) is immediate. It corresponds to $g = 0$ or $\tan 2\theta_{k_x}$ = 0, a situation when the system assumes a paramagnetic order at $\theta_{k_x} = 0$.

An easy analytical solution of Eq.~(\ref{eq:gap3}) is obtained if one assumes a constant
Coulomb interaction $2\pi e^2/|\mb k-\mb k'| = \gamma$, such that the right hand side does not depend on $k_x$ anymore and the gap is a constant. Following Ref.~\cite{awo62}, the integration domain can be chosen to be a rectangle in $\mb k$ centered at $0$,of lengths $L_x$ and $L_z$ that incorporate the states that are likely to be distorted by the gap formation in Fig.~\ref{fig1}. Then, the gap equation can be integrated directly and the gap is obtained as,
\be
g = \frac{L_x\hbar^2Q}{m^*\sinh \left(\frac{4\hbar^2Q\pi^2}{m^*\gamma L_y}\right)}\;. \label{eq:gap-sol}
\ee
Although, in this approximation, $g$ does not correspond to a ground state, since the exchange corrections to the single-particle energies are neglected, its variation with the system parameters is instructive for more exact calculations \cite{marinescu-prep}.

Finally, the fractional spin-polarization associated with states $f_{\mb k,-}=1$ is obtained to be, from Eqs.~(\ref{eq:states}),
\be
\mb P =\sum_{\mb k}\langle \Psi_{1\mb k}| \mb \sigma |\Psi_{1\mb k}\rangle  = \hat{\mb x} \sum_{\mb k} \sin 2\theta_{\mb k} + \hat{\mb z} \sum_{\mb k}\cos 2\theta_{\mb k}\;.
\ee
indicating that the polarization maintains a constant direction in space. Further, since the $\hat{\mb z}$ component is also zero, on account of the oddness of $\cos 2\theta_{k_x}$, the direction of the polarization is parallel to that of the displacement vector $\mb Q$.

In conclusion, in a quantum well with equal Rashba-Dresselhaus spin-orbit couplings, the ground state of the electron system is likely an itinerant antiferromagnet at low temperatures, with a fractional polarization oriented along the $\hat{\mb x}$ axis, the direction in which the single electron spectrum is displaced. It is important to point out that this spin arrangement might be itself unstable in respect with a paramagnetic transition depending on system parameters, such as particle density and the strength of the coupling $\alpha$.

This research was  supported in part by the National Science Foundation under Grant No. PHY11-25915.
\end{document}